\documentclass[baaa]{baaa}

\usepackage[pdftex]{hyperref}
\usepackage{subfigure}
\usepackage{natbib}
\usepackage{helvet,soul}
\usepackage[font=small]{caption}

\contriblanguage{1}

\contribtype{1}

\thematicarea{7}

\title{Kinematics of the Local Group gas and galaxies\\ in the {\sc Hestia} simulations}

\titlerunning{Kinematics of the Local Group in the Hestia simulations}

\author{
L. Biaus\inst{1,2},
S.E.~Nuza\inst{2,3} \&
C.~Scannapieco\inst{1,2}
}

\authorrunning{Biaus et al.}

\contact{lbiaus@df.uba.ar}

\institute{
Departamento de Física, Facultad de Ciencias Exactas y Naturales, UBA, Argentina
\and
Consejo Nacional de Investigaciones Cient\'{\i}ficas y T\'ecnicas, Argentina
\and
Instituto de Astronom{\'\i}a y F{\'\i}sica del Espacio, CONICET--UBA, Argentina
}

\resumen{El Grupo Local (LG) está formado por dos galaxias espirales gigantes, la Vía Láctea (MW) y Andrómeda (M31), y varias galaxias más pequeñas. MW y M31 se están acercando a una velocidad radial de $-109\,$km\,s$^{-1}$. Evidencia observacional sugiere que hay un movimiento general de caída de gas y galaxias en el LG, dominado por la dinámica de sus dos miembros principales. Desde nuestra perspectiva, este flujo imprime un patrón dipolar de velocidades en el cielo cuando se remueve la rotación Galáctica. Investigamos las propiedades cinemáticas del gas y galaxias en el LG usando un conjunto de simulaciones de alta resolución de la colaboración {\sc Hestia} (High-resolution Environmental Simulations of The Immediate Area). Nuestras simulaciones incluyen la cosmografía correcta alrededor de regiones similares al LG
. 
Construimos mapas del cielo desde el sistema de reposo local, Galáctico y del LG. Encontramos que, al remover la rotación Galáctica, se establece un dipolo de velocidad radial en la dirección preferencial correspondiente al baricentro del LG para material por fuera del radio virial de la MW cuando la velocidad radial relativa entre la MW y M31 es similar al valor observado. Este resultado favorece un escenario donde el gas y las galaxias caen hacia el baricentro del LG, produciendo el dipolo de velocidades observado.}

\abstract{The Local Group (LG) consists of two giant spiral galaxies, the Milky Way (MW) and Andromeda (M31), and several smaller galaxies. The MW and M31 are approaching each other at a radial velocity of about $-109\,$km\,s$^{-1}$. Observational evidence suggests that there is an overall infalling motion of gas and galaxies in the LG, dominated by the dynamics of its two main members. From our perspective, this flow imprints a velocity dipole pattern in the sky when Galactic rotation is removed. We investigate the kinematic properties of gas and galaxies in the LG using a suite of high-resolution simulations performed by the {\sc Hestia} (High-resolution Environmental Simulations of The Immediate Area) collaboration. Our simulations include the correct cosmography surrounding LG-like regions
. 
We build sky maps from the local, Galactic and LG standard of rest reference frames. Our findings show that the establishment of a radial velocity dipole near the preferred barycentre direction is a natural outcome of simulation kinematics for material \textit{outside} the MW virial radius after removing galaxy rotation when the relative radial velocity of MW and M31 is similar to the observed value. These results favour a scenario where gas and galaxies stream towards the LG barycentre, producing the observed velocity dipole.}


\begin{document}

\maketitle

\section{Introduction}\label{S_intro}

The Local Group (LG) encompases the Milky Way (MW), Andromeda (M31) and several other minor galaxies. The MW and M31 are on a collision course, due to the general motion of LG galaxies towards the group’s barycenter \citep[e.g,][]{BT08}. Observations suggest that a giant multiphase gas halo surrounds the MW and M31 
and possibly point out to the existence of LG gas located outside the virial radius of the MW.

Observational evidence for the kinematics of the LG gas is mainly derived from absorption-line measurements in the spectra of background sources, probing the chemical composition of intervening material by studying the imprint of a variety of ions at different wavelengths. In particular, \cite{Richter17} analysed a large sample of high-velocity absorbers drawn from archival UV spectra of extragalactic background sources and determined the existence of a velocity dipole at high Galactic latitudes (as seen from the Local Standard of Rest or LSR). They interpreted this as possible evidence for intragroup gas streaming towards the LG barycenter as a result of the expected general flow of gas and galaxies inwards. In this work, we studied if this interpretation is consistent with the simulated kinematics of gas in LG-like regions belonging to the  {\sc Hestia} cosmological suite of simulations.

This proceeding is organised as follows. In Sec.~\ref{s:sims} we introduce the main aspects about the LG simulations used in this work. In Sec.~\ref{s:analysis} we describe how we set up the Sun's position and velocity within the simulated MW to define the velocity reference frames used throughout this work. In Sec.~\ref{s:results} we present the predictions for gas kinematics in the analysed simulations. Finally, in Sec.~\ref{s:conclusions} we summarise and discuss our results.

\section{Simulations}\label{s:sims}

Simulations in the {\sc Hestia} project aim at obtaining galaxy systems resembling the LG within a cosmological context. Here we present a summary of its main characteristics, and for further details we refer the reader to \cite{Libeskind20} and references therein.

The simulations were run using the moving-mesh, cosmological code {\sc Arepo} \citep{Springel10, Weinberger20}, which computes the joint evolution of gas, stars and dark matter (DM) by solving the gravitational and ideal magnetohydrodynamics (MHD) equations coupled to the {\sc Auriga} galaxy formation model \citep{Pakmor13, Grand17}.

In this work, we use three simulation runs, dubbed as $37\_11$, $9\_18$ and $17\_11$, consisting of two overlapping high-resolution spheres of $2.5\,h^{-1}\,$Mpc radius centred on the MW and M31 candidates that are surrounded by lower-resolution particles. All simulations in the {\sc Hestia} project aim to reproduce the following main cosmographic features: the Virgo cluster, the local void and the local filament. Throughout this proceeding, we will focus on realisation $17\_11$, as it has the most similar infall velocity between the two main haloes to the observed value (infall velocites are $+9$, $-74$ and $-102$~km s$^{-1}$ for realisations $37\_11$, $9\_18$ and $17\_11$ respectively, whereas the observed value reported by \citealt{vanderMarel12} is $109 \pm 4.4$~km s$^{-1}$). For further details regarding the analysis of  realisations $37\_11$ and $9\_18$, we refer the reader to \cite{Biaus22}.

\section{Analysis}\label{s:analysis} 

\begin{figure}
    \centering
    \includegraphics[width=\columnwidth]{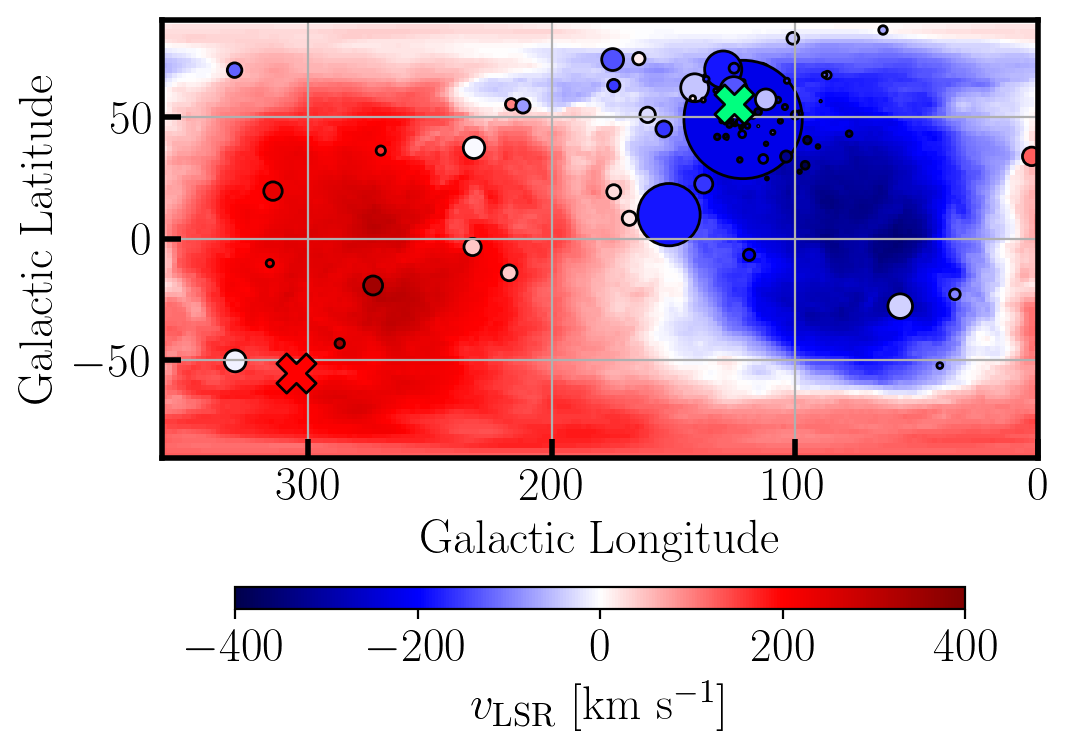}\vspace{-0.2cm}
    \caption{Sky distribution of galaxies with $L_V > 10^4\, L_{V_\odot}$ {\it outside} the virial radius of the MW analog for realisation $17\_11$. Solid circles are colour-coded by LSR velocity with radius indicating a size of 2$R_{200}$ for each halo. The mean mass-weighted gas velocity at any given line of sight is also shown. Crosses indicate the LG barycentre (green) and anti-barycentre (red) directions.}
    \label{fig:LSR_sky}
\end{figure}

In this work, we refer to the Local Standard of Rest (LSR), the Galactic Standard of Rest (GSR) and the Local Group Standard of Rest (LGSR). To define these frames of reference, we locate the Sun in the simulated MW's midplane at a distance of $8$~kpc from the galactic centre and at an azimuthal angle from which the longitude of the simulated M31 matches the observed one.

When considering the motion of the Sun around the MW, we take the galaxy's circular velocity at the observer radius pointing towards $(l,b)=(90^{\circ},0^{\circ})$ as the Sun's velocity vector ($|\mathbf{v}_{\odot}| = 218$~km\,s$^{-1}$ for realisation $17\_11$). This defines our LSR reference frame. When referring to the GSR, we simply exclude the velocity field produced by the rotation of the galaxy  while, to refer to the LGSR, we additionally exclude the radial motion of the MW with respect to the LG barycentre \citep[e.g.][]{Karachentsev96}

\section{Results}
\label{s:results}
\subsection{Simulated LSR map}

\begin{figure}
    \centering\includegraphics[width=\columnwidth]{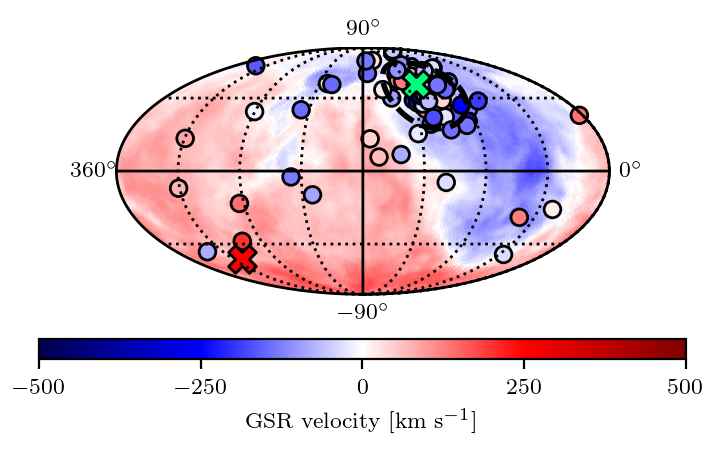}
    \caption{Mollweide projection of mean mass-weighted gas and galaxies GSR velocities in realisation $17\_11$ \textit{outside} the MW's virial radius. Crosses indicate the LG barycentre (green) and anti-barycentre (red) directions. The black dotted line outlines the circumgalactic medium of M31.}
    \label{fig:GSR}
\end{figure}

Fig.~\ref{fig:LSR_sky} shows the sky distribution of galaxies viewed from the LSR (colour-coded by radial velocity) that are {\it outside} the MW's virial radius in the realisation $17\_11$ with a lower cutoff of $L_V\sim 10^4\,L_{\odot}$ (symbols sizes are proportional to halo apparent size, based on its $R_{200}$ and distance from the observer). The coloured background is a map for LSR gas velocity, which we compute by performing a mass-weighted average of all gas cells along the line of sight, for gas outside the MW's virial radius up to $1000$~kpc.

The distribution of gas radial velocity displays a perceptible velocity dipole pattern, resulting from the combination of the MW's galaxy rotation and its motion towards the LG barycentre. Not surprisingly, galactic rotation is the main responsible for the sharp velocity contrast seen between $0^{\circ} < l < 180^{\circ}$ and $180^{\circ} < l < 360^{\circ}$. This effect is, however, more relevant at latitudes close to the galactic plane. At higher latitudes, the relative motion between MW and M31 also imprints a dipole pattern in the sky whose strength and overall sign depends on the absolute velocity between the two main galaxies. The map shows a velocity dipole for the gas that persists even at high latitudes.

\begin{figure*}
    \centering
	\includegraphics[width=1.8\columnwidth]{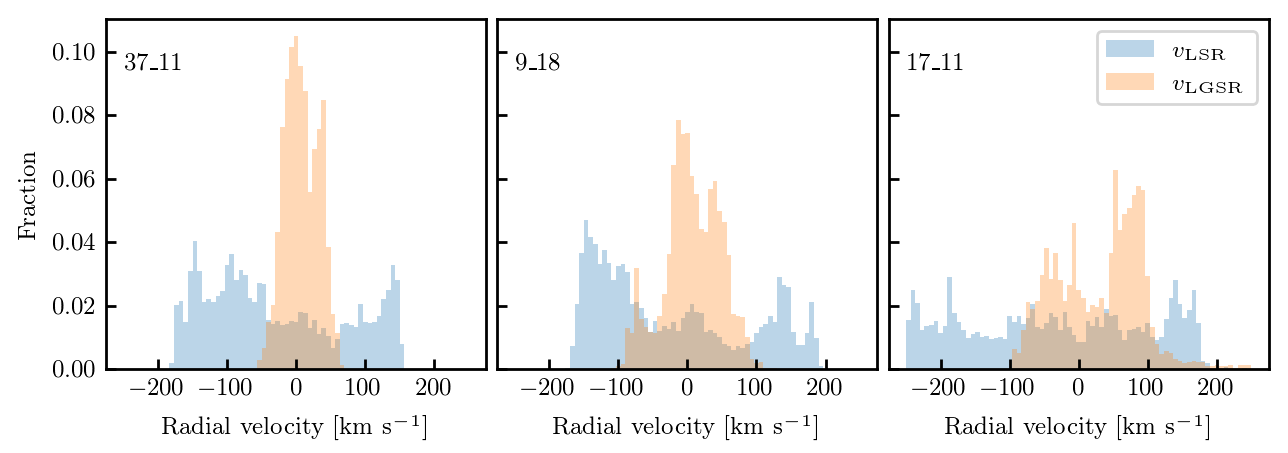}

    \caption{LSR and LGSR velocity distributions for gas outside the MW virial radius for our three high-resolution realisations. We probe 6080 directions evenly distributed along the quadrants containing the barycentre and anti-barycentre in each simulation.
    }
    \label{fig:LGSR_histo}
\end{figure*}

\subsection{Excluding Galactic rotation}

In Fig. \ref{fig:GSR}, we show the GSR velocity sky map for the gas outside the simulated MW's virial radius. Overplotted dots are galaxies belonging to the LG, where the green (red) cross indicates the sky position of the barycentre (anti-barycentre) and the black dotted line outlines the virial radius of M31. As it can be clearly seen in this figure, a velocity dipole in the general barycentre-antibarycentre direction is evident, both for gas and galaxies, but particularly for the former. The approaching gas (blue) extends far beyond M31's circumgalactic medium, which supports the idea that this velocity dipole is related not only to the relative motion of the MW and M31 but rather to the global kinematics of the LG, implying that some of the observed absorption lines towards the approximate direction of M31 could be linked to LG gas.

\subsection{Moving to the LGSR frame}

Observationally, a necessary (but not sufficient) condition for the gas to be located outside the virial radius of the MW 
is to observe a decrease in the spread of the radial velocity distributions as one moves from the LSR to the GSR and LGSR reference frames \citep{Sembach03}. With this in mind we plot, in Fig.~\ref{fig:LGSR_histo}, the radial velocity distribution of gas in the {\it general} barycentre and anti-barycentre directions in our three simulations. For comparison, we show distributions for both LSR and LGSR reference frames. The histograms are done selecting all lines of sight found at two sky regions separated by the lines $l=180^{\circ}$ and $b=0^{\circ}$ (i.e., quadrants {\sc II} and {\sc IV} in simulation $9\_18$, and {\sc I} and {\sc III} in simulations $37\_11$ and $17\_11$, for the barycentre and anti-barycentre ``regions'', respectively). The figure shows that after transforming the gas velocities to the LGSR, the standard deviation of the distributions noticeably decreases. Moreover, we find that for gas outside MW's virial radius, the LGSR velocity distribution is bimodal in the three simulations, strengthening the usual interpretation made by observers of LG gas flowing towards the barycentre for material seen towards these sky regions \citep[see e.g.][and references therein]{Bouma19}.


\section{Conclusions}
\label{s:conclusions}

In this work, we have studied the kinematic properties of LG gas and galaxies in a suite of three high-resolution ($37\_11$, $9\_18$ and $17\_11$) re-simulations of the LG belonging to the {\sc Hestia} project \citep{Libeskind20} as seen from an observer located at the Sun's position.

We have shown that the existence of a radial velocity dipole for both gas and galaxies outside the MW virial radius from the LSR is a natural outcome in realisation $17\_11$ where the approaching velocity of MW and M31 galaxy candidates is similar to the observed value. We also show that, after removing the MW's galactic rotation, a velocity dipole persists. 


When studying gas kinematics in the LGSR, the radial velocity distributions of gas outside the virial radius towards the barycentre/anti-barycentre quadrants in all simulations are in line with the results of \cite{Richter17}, suggesting that the usual observational interpretation of gas and galaxies flowing towards the LG barycentre may be justified. 

\begin{acknowledgement}
The authors acknowledge support provided by UBACyT 20020170100129BA.
\end{acknowledgement}

\bibliographystyle{baaa}
\small
\bibliography{bibliografia}
 
\end{document}